\documentclass{IOS-Book-Article}

\usepackage{mathptmx}
\usepackage{url}
\usepackage{hyperref}

\begin{document}
\begin{frontmatter}              

\title{Knowledge Affordances for \\Hybrid Human-AI Information Seeking}

\author{\fnms{Irene} \snm{Celino}
\thanks{Accepted at Hybrid Human Artificial Intelligence Conference (HHAI 2026) -- \copyright ~Irene Celino, 2026}
}

\runningauthor{Irene Celino}
\address{Cefriel -- viale Sarca 226, 20126 Milano, Italy \\E-mail: \href{mailto:irene.celino@cefriel.com}{irene.celino@cefriel.com}}

\begin{abstract}
As information ecosystems grow more heterogeneous, both humans and artificial agents increasingly face a simple yet unresolved question: when seeking knowledge, whom should we ask, and why? Inspired by how people intuitively "read a room", this paper introduces the concept of knowledge affordance (KA)  to systematize how agents identify meaningful opportunities for information seeking in hybrid human–AI environments.

Rather than introducing a fully formed framework, we propose KAs as declarative, semantically grounded descriptions of what a knowledge source can offer, for which kinds of questions, and with which contextual properties. Additionally, we suggest that KAs are relational, possibly emerging from the interplay between the agent's task, preferences and situational factors.

Our contribution is thus a conceptual proposal that connects different research streams, including affordances, semantic web services, knowledge engineering and querying, and mutual intelligibility. 
We sketch possible research directions to build KA‑aware systems that navigate information spaces with greater transparency, adaptability and shared understanding.
\end{abstract}

\begin{keyword}
knowledge affordance \sep 
agent interaction \sep
information seeking 
\end{keyword}
\end{frontmatter}

\thispagestyle{empty}
\pagestyle{empty}

\section{Introduction}

Imagine walking past a large room from which you hear background music and the sound of people chatting. As you approach the entrance, you realize that the room is crowded: some people stand alone, others are engaged in small conversations, while a few gather around a corner where food and drinks are available. The atmosphere is inviting, and you feel inclined to step in and understand what is going on, possibly to take part in it. However, you do not recognize anyone in the room. To get oriented, you need to ask a stranger for information: who do you choose to approach?

Different people would make different choices in this situation: the closest person, to minimize effort; someone with a friendly appearance, as a signal of social acceptance; a lively group, to expand own social network; a person standing alone, assuming that this would cause the least disturbance; someone perceived as more similar or compatible (e.g., similar age, comparable physical characteristics, or familiar style) assuming that shared traits would make interaction easier and communication more effective.

These choices are not random. The perceived opportunity to ask for information -- the \emph{affordance} of doing so -- depends on multiple factors: characteristics of the \emph{environment}, properties and goals of the requester \emph{agent}, and relational aspects emerging from their \emph{interaction}, such as context, intentions, and social norms.

A closely related challenge arises when the subject seeking information is not a human, but an AI agent operating (possibly on behalf of a human user) within a ecosystem of heterogeneous and potentially unfamiliar knowledge sources. In hybrid environments where humans and AI systems coexist and interact, identifying which sources are relevant, accessible, or appropriate to consult becomes a non-trivial problem. 

This paper introduces the concept of \emph{knowledge affordance} to characterize and reason about the conditions under which knowledge sources can be perceived as actionable by different agents. By framing information seeking as an affordance-driven process, we aim to provide a conceptual tool to support more effective interaction with complex knowledge ecosystems involving both human and artificial agents.

\section{Motivation}

The motivation for introducing a new conceptual framework stems from the profound transformation that information seeking has undergone over time within information systems. Early approaches were primarily centered on \emph{database querying}: applications provided access to information through interfaces that exposed the results of structured formal queries, often authored by expert users with specific skills. With the advent of search engines, information seeking shifted toward \emph{keyword-based search}, relying on purpose-built indexes to match user queries with relevant documents. More recently, the widespread adoption of generative AI and large language models (LLMs) has further changed user interaction paradigm, now commonly based on \emph{natural-language question answering}, where AI agents interpret inquiries and provide responses~\cite{hogan2025large}.

Despite this evolution, a formal model is still missing to capture how information seeking unfolds in a situated manner, namely, how agents decide what to ask, whom to ask, when, and for what purpose. This limitation becomes particularly evident in \emph{rich and diverse ecosystems that combine heterogeneous information sources}, such as services, knowledge graphs, and AI agents, each characterized by different access mechanisms, representational assumptions, and capabilities.

Moreover, information seeking in hybrid human-AI systems is inherently \emph{collaborative}. It requires continuous adaptation and coordination among multiple agents, both human and artificial, which must iteratively interact to establish mutual understanding and jointly achieve a desired outcome. In such settings, information seeking cannot be reduced to a single query-response interaction, but rather emerges as a dynamic process shaped by goals, context, and interaction history.

Finally, most existing information retrieval and question answering models implicitly assume that the agent already knows where to look for information. In today’s digital landscape, however, agents operate within ecosystems of heterogeneous and evolving \emph{knowledge sources}, which differ in availability, scope, and capabilities. As a result, deciding which sources can be meaningfully accessed and exploited becomes a central challenge that is not adequately addressed by current information-seeking models.

\section{Background and Conceptual Foundations}

In this section, we recap the main conceptual foundations used to define the concept of knowledge affordance, rather than providing a comprehensive literature review.

\subsection{Affordance Theory}

The concept of \emph{affordance} was originally introduced in ecological psychology to describe what an environment or an artifact offers to an actor, emerging from the relationship between the actor and the environment rather than residing exclusively in either of them~\cite{gibson1978ecological}. An affordance thus refers to an \emph{action possibility}: a potential action that becomes available depending on the actor’s capabilities, goals, and perceptions, as well as on the material and symbolic properties of the environment or artifact. In this sense, affordances are inherently relational and context-dependent, as they capture how certain actions are enabled or constrained by the specific fit between actor and environment.

In design theory, Don Norman adapted the notion of affordance to focus on how the perceived properties of artifacts suggest possible uses to users, influencing usability and interaction~\cite{norman2013design}. This interpretation has been widely adopted in human-computer interaction (HCI), where affordances are used to understand about how interface elements support, guide, or limit user actions within socio-technical settings. The concept has also been explored in robotics, where affordances are modeled  to allow robots to recognize actionable properties of objects and use them to guide task execution~\cite{jamone2016affordances}.

In this work, we build on the idea of affordance as an action possibility and apply it to the specific case of information and knowledge seeking. We use this theory to identify when and how knowledge sources can be perceived as actionable by an agent. 

\subsection{Semantic Description of Services}

Semantic Web Services (SWS)~\cite{fensel2011sws} introduced ontology-based frameworks to provide machine-processable descriptions of service capabilities. Their core idea was to use formal semantic descriptions to enable automated reasoning about what a service does (\emph{functional aspects}) and under which conditions it can be effectively used (\emph{non-functional aspects}), with the goal of supporting tasks such as service discovery, selection, and composition~\cite{martin2007owls,roman2005wsmo,dellavalle2005discovery}.

Early SWS approaches, most notably OWL-S and WSMO, relied on explicit semantic annotations to describe service behavior, requirements, and quality-related properties. In particular, WSMO explicitly introduced the notion of goals and mediators to address interoperability, while both frameworks emphasized the role of non-functional properties, such as quality of service or user preferences, in service matching and selection~\cite{panziera2010nfp}.

Despite their conceptual strength, SWS approaches did not achieve widespread adoption, mainly due to their complexity, limited tooling, and the effort required to produce  semantic descriptions. At the same time, the evolution toward lightweight service technologies like REST led to the current predominantly syntactic interface specifications, such as OpenAPI~\cite{openapi}, which are easier to adopt but lack machine-interpretable semantics.

More recently, the growing availability of AI models and agents has renewed interest in describing and exposing capabilities in a machine-readable form. On the one hand, initiatives such as Model Cards~\cite{mitchell2019modelcards} aim to document the characteristics and intended use of trained machine learning models. On the other hand, emerging protocols for agent interoperability, such as MCP, focus on exposing functions and interfaces that AI agents can invoke, effectively treating them as actionable capabilities.

We draw inspiration from these lines of work to argue that machine-readable descriptions should be leveraged also in the domain of information and knowledge seeking. In particular, we build on the distinction between functional and non-functional properties to motivate a richer description of knowledge affordances, capturing both what knowledge sources offer, and under which conditions they can be effectively used.

While these approaches provide expressive formalisms to describe \textit{what} a service or source can do, they offer limited support for modeling the \textit{situated decision} of whether, when, and why a given source should be consulted by a specific agent in context. We do not aim to replace formalisms such as OWL‑S, but to complement them by making the appropriateness and activation of a knowledge source an explicit, first-class concern. We focus on the relational and context-dependent activation of knowledge sources, including human agents, rather than on the static semantic description of service interfaces alone.

\subsection{Knowledge Querying and Hybrid System Design}

In ontology engineering, \emph{competency questions} (CQs) are commonly used to capture the information needs that an ontology is expected to support~\cite{alharbi2024review}. A CQ expresses a typical user question over a conceptual or domain model and, when a knowledge graph adopts a given ontology, CQs can be seen as abstract templates of the kinds of questions that users may wish to ask over that knowledge graph.

Question answering over knowledge graphs is a long-standing research area, spanning from early ontology-driven approaches~\cite{lopez2012poweraqua}, to more recent methods using LLMs to translate natural language questions into SPARQL queries~\cite{hernandez2025fewshotsparql}. Community-driven challenges also aim to establish shared evaluation settings to compare such approaches~\cite{TEXT2SPARQL2025}.

A related line of research concerns the design of hybrid systems that combine learning and reasoning components~\cite{vanharmelen2019boxology,vanbekkum2021modular} or neuro-symbolic approaches~\cite{deboer2025patternllmnesy}. Those works propose reusable \emph{design patterns} to structure the interaction and information flow between heterogeneous components. Similarly, \emph{knowledge engineering} methodologies have been investigated for their ability to support collaboration and coordination in hybrid human-AI systems~\cite{tiddi2023ke4hi}.

Taken together, these contributions provide us with essential building blocks to make explicit \emph{when} and \emph{why} a given knowledge source should be considered actionable by an agent in a specific context: CQs help characterize information needs, question answering techniques enable the transformation of user queries into actionable requests, and design patterns and knowledge engineering methodologies guide the integration of heterogeneous systems. 

\subsection{Hybrid Human–AI Intelligence \& Mutual Intelligibility}

In hybrid human-AI systems, \emph{mutual intelligibility}~\cite{mestha2024mutualintel} is a property of human-AI cooperation through which both agents are able to make their states, goals, and reasoning processes intelligible to each other in an interactive and bidirectional way. This mutual understanding supports the alignment of meanings and assumptions and enables joint planning and coordinated action. Mutual intelligibility goes beyond pure machine-to-human explainability and beyond trust alone, as it requires explicit mechanisms that allow both parties to actively understand and adapt to each other.

A formal protocol to design and measure mutual intelligibility is proposed in~\cite{mestha2024mutualintel}, which explicitly distinguishes between one-way and two-way intelligibility. Along similar lines, works inspired by Theory of Mind~\cite{wang2022mutualtom} model human-AI communication as a reciprocal process in which the interacting parties infer each other's intentions and mental states. Mutual intelligibility has also been identified as a key requirement for cooperation and transparency in hybrid human-AI systems~\cite{shi2022survey}. Finally, with specific reference to neuro-symbolic AI and knowledge graphs, mutual understanding has been defined as the ability to share, exchange, and govern knowledge between humans and information systems~\cite{celino2025mutualunder}, and can be operationalized through mechanisms supporting mutual intelligibility in collaborative settings.

In this work, we argue that knowledge affordances can contribute to mutual intelligibility by providing a semantic interface that supports collaboration and interaction in information-seeking tasks in hybrid human-AI systems. 

\section{Concept of Knowledge Affordance}

Building on the related work discussed so far, we observe that existing approaches provide powerful mechanisms to describe knowledge sources, query structured information, and orchestrate hybrid human-AI systems. However, they offer limited support for reasoning about when and why a given knowledge source should be considered actionable by an agent in a specific situation. To address this gap, we introduce the notion of knowledge affordance.

\subsection{Definition and characteristics}

A \emph{knowledge affordance} (KA) is an explicit semantic interface that characterizes what a knowledge source can offer to an agent in the context of an information-seeking task. It describes which types of questions a source can support, and under which constraints. Knowledge sources may include knowledge graphs, services, tools, or other human or artificial agents. By making these aspects explicit, a KA enables an agent -- human or artificial -- to reason about how to exploit available sources to achieve its information-seeking goals.

The output of a KA is not a direct answer to a query. Instead, a KA supports the construction of an \emph{interrogation plan}, i.e., an explicit, structured sequence of information-seeking actions that specifies how different knowledge sources can be selected, combined, and queried. Each step captures what to ask, to whom, and under which constraints. Such a plan may include question templates inspired by competency questions, routing or prioritization strategies based on contextual or agent-specific preferences, and explicit motivations for source selection, supporting explainability. In this sense, KAs operate at the level of planning and coordination, rather than execution.

Consistently with the original notion of affordance, KAs have a fundamentally \emph{relational nature}. They are not static properties of a knowledge source, but emerge from the interaction between several elements: the characteristics of the source itself, which can be interpreted in terms of capabilities and non-functional properties; the current task and information need; the requesting agent, with its state, preferences, and constraints; and the operational context in which the interaction takes place. This perspective closely reflects the psychological interpretation of affordances as action possibilities, arising from the fit between an actor and its environment.

From an engineering perspective, we conceptualize a KA as a \emph{semantic interface} that is both explicit and situated. The \emph{explicit} component captures machine-interpretable descriptions of knowledge sources, including capability descriptions inspired by semantic web services, classes of competency questions, the scope of coverage over a knowledge graph (e.g., entities, subgraphs, or schemas), grounding to concrete services, endpoints, or agents, and a set of non-functional properties. These properties may include computational cost, latency or reachability, trust and reliability, or contractual and policy-related conditions, as well as human factors such as perceived intrusiveness, social effort, or appropriateness in a given organizational or social context.

At the same time, the activation of a KA is inherently \emph{situated}. 
It depends on agent-specific factors, such as preferences, cognitive constraints, and strategic goals, which in hybrid settings often reflect proxy representations of human values, social norms, or interaction constraints. 
Some non-functional properties capture dimensions that are analogous to social cues in human interaction, such as how easy a source is to engage with or how demanding its use may be in a given context. While these properties can be explicitly described, their relevance and activation emerge dynamically from the interaction between the agent, the task, and the environment.

\subsection{Formal Model for Knowledge Affordances}

To make the notion of knowledge affordance operational while keeping it lightweight, we propose a minimal formal model that distinguishes between (i) the explicit semantic description of a knowledge source and (ii) the situated activation of that description by a specific requester in a given context.

A \emph{knowledge affordance} KA for a knowledge source is modeled as a tuple:
\begin{equation}
\mathrm{KA} = \langle\: C,\: CQ,\: S,\: NFP,\: G \:\rangle
\label{eq:ka}
\end{equation}
where $C$ denotes the capabilities offered by the knowledge source, $CQ$ the classes of competency questions it can address, $S$ the relevant content scope (e.g., over a knowledge graph), $NFP$ a set of non-functional properties influencing its use, and $G$ the grounding to concrete endpoints, services, tools, or agents that implement those capabilities.

Intuitively, this tuple captures \emph{what} a source can do, \emph{for which} kinds of questions, \emph{over which} knowledge, \emph{under which} qualitative or quantitative conditions, and \emph{how} it can be accessed. The model is intentionally agnostic with respect to representation choices: each element may be encoded using ontologies, metadata vocabularies, or other semantic descriptions, depending on the application setting.

KAs are activated by a \emph{requester}, modeled as:
\begin{equation}
R = \langle\: A,\: T,\: X \:\rangle
\label{eq:requester}
\end{equation}
where A represents the requesting agent (including preferences, constraints, and interaction style, which in hybrid systems may act as an explicit proxy for a human user’s goals, values, and situational judgments), $T$ is the current information-seeking task (possibly represented at an abstract or partial level and not necessarily fully disclosed to all knowledge sources), and $X$ denotes the operational context (e.g., conversational state, availability of sources, or environmental conditions).

Given a set of available KAs, an agent does not use all of them indiscriminately, but discovers and selects those that are appropriate for its current state. We capture this through a \emph{selection} operator:
\begin{equation}
\textsc{SelectKA}(R) = \arg\max_{\mathrm{KA}_i} \;\; \textsc{Appropriateness}(\mathrm{KA}_i,\: R).
\label{eq:selection}
\end{equation}
The appropriateness function combines, in a context-dependent manner, the alignment between aspects of the task $T$ and the supported competency-questions $CQ$, the compatibility between required and offered capabilities $C$, the relevance of the scope $S$, the match between non-functional properties $NFP$ and agent preferences $A$, and the feasibility of using the affordance grounding $G$ in context $X$. Different instantiations are possible, but the key point is that selection is relational and situated.

When activated, a KA does not directly return an answer. Instead, it supports the construction of an \emph{interrogation plan}. 
Each step in the plan specifies an actionable interaction with one or more grounded knowledge sources, such as issuing a query to a knowledge graph endpoint, invoking a service or tool, or delegating a subtask to another agent. The plan may be accompanied by query or prompt templates derived from the relevant competency-questions, as well as routing or prioritization strategies that reflect agent preferences or contextual constraints. 

The selection process can also provide an \emph{explanation} for a KA activation, making the rationale behind source selection explicit and available for human inspection, acceptance, or rejection. 
This explicitly distinguishes KAs from question answering or retrieval-augmented generation approaches, but also supports human oversight and control over the information-seeking process~\cite{TrustworthyAI}. 
This makes the planning process transparent and supports mutual intelligibility in hybrid human-AI settings.

\subsection{Illustrative Example}

Consider a large multinational organization that must take a strategic decision regarding compliance with regulatory frameworks such as the AI Act or the Data Act. The information-seeking task requires assessing implications across legal, technical, and organizational dimensions under conditions of uncertainty. In this setting, each source corresponds to a distinct $\mathrm{KA} = \langle\: C,\: CQ,\: S,\: NFP,\: G \:\rangle$, while the organizational goal instantiates the requester $R = \langle\: A,\: T,\: X \:\rangle$ used to construct the interrogation plan.

The organization can draw on multiple heterogeneous knowledge sources. These include internal compliance documents and legal opinions (authoritative, but potentially outdated); external analyses and benchmarks (timely, but weakly contextualized); AI-based tools that summarize and compare regulatory requirements (fast, but opaque and possibly biased); and human experts, such as legal or compliance officers (context-sensitive, but costly and limited in availability). Each source exposes a KA describing the types of questions it can address, its scope, and non-functional properties such as reliability, interpretability, update frequency, or effort required for engagement.

Given the task, an agent evaluates multiple KAs and constructs an interrogation plan. For example, it may combine AI-based tools to obtain an initial overview, internal documentation to assess compliance gaps, and human experts to validate interpretations or resolve ambiguities. KA selection involves different trade-offs: sources with broader coverage or efficiency may be deprioritized in favor of those perceived as more trustworthy, or aligned with organizational norms in a high-stakes decision. By making these trade-offs explicit, the agent enables a human decision-maker to inspect the explanation behind the proposed plan and ultimately accept, modify, or override it based on risk, accountability, or other human-centered considerations.

\section{Research Directions}

In this paper, we introduced the concept of knowledge affordance and argued for its relevance in hybrid human-AI information-seeking settings. We consider this contribution as a first step toward a broader research agenda, which we briefly outline in the following.

A first set of research directions concerns \emph{foundational aspects} of the KA concept and its concrete specification. This includes investigating how non-functional properties of KAs can be defined, represented, and compared in a principled way, while preserving their relational and context-dependent nature. Similarly, modeling agent preferences, constraints, and requests in a form that supports interoperability across different systems and ecosystems remains an open challenge. Finally, different strategies for assessing the appropriateness of a KA for a given request deserve further investigation, as alternative discovery, ranking, and selection mechanisms may lead to different trade-offs.

A second set of directions relates to \emph{methodological and engineering challenges}. Defining and maintaining the explicit semantic interface of a KA raises questions about suitable modeling, design patterns, and tool support. In this respect, knowledge engineering methodologies can play a central role, both to leverage existing artifacts, such as ontologies and competency questions, and to support the systematic construction and evolution of KAs. In addition, planning interrogation strategies and orchestrating multiple sources or agents to satisfy an information-seeking goal opens the door to the application of different planning and optimization techniques.

\emph{Explainability and evaluation} constitute another important research direction. Making the selection of a specific KA transparent is closely related to existing work in explainable~\cite{hoffman2019xaimetrics,kim2024xaihumaneval} and neuro-symbolic AI~\cite{renkoff2024nesyeval}, but shifts the focus from explaining answers to explaining source selection and coordination. At the same time, evaluating the effectiveness of KAs should go beyond task-level success, by also considering how well a KA supports adaptation over time, reuse across contexts, or the progressive alignment between agents and knowledge sources.

Finally, with specific reference to \emph{hybrid human-AI systems}, KAs raise questions about how human preferences, social constraints, and shared understanding can be explicitly modeled and taken into account. Also, hidden or even unconscious human beliefs and assumptions could be difficult to reveal or make explicit, adding an additional challenge.  Addressing these aspects is essential to support mutual intelligibility and effective collaboration, especially in multi-agent ecosystems where information seeking is a collective and evolving process.

\section{Conclusions}

We started this paper with the example of a room full of strangers, where different strategies can be adopted to decide whom to approach for information. Similarly, in complex information ecosystems, both human and AI agents need support not only to ask questions, but also to orient themselves among heterogeneous knowledge sources.

In this work, we introduced the concept of \emph{knowledge affordance} (KA) to address this need. We framed KAs as explicit semantic interfaces that make it possible to reason about which sources to use, for which kinds of information needs, under which conditions, and with what implications. We also proposed a lightweight formal model to capture the relational and situated nature of KAs, highlighting how source selection and coordination can be treated as first-class objects. 

More broadly, KAs point toward a view of information seeking in which agents, human or artificial, navigate knowledge ecosystems in a transparent and intelligible way, choosing appropriate interlocutors and strategies based on context and preferences. We argue that this capability is essential for hybrid human-AI systems, and that KAs provide a conceptual foundation to support it.

\vspace{0.2cm}\noindent\emph{Acknowledgements} This research has been partially supported by the PERKS project, co-funded by the European Commission under the Horizon Europe programme, Grant Agreement no. 101120323.

\bibliographystyle{vancouver}
\bibliography{biblio}

\end{document}